\documentclass[intlimits,twoside,a4paper]{article}

\usepackage{graphicx}
\usepackage{amssymb, amsmath}
\usepackage[T2A]{fontenc}
\usepackage[cp1251]{inputenc}

\usepackage{cmpj}
\issue{2011}{14}{1}{13601}

\doinumber{10.5488/CMP.14.13601}

\title[Core-softened model fluid]
{Microscopic structure and thermodynamics of a core-softened model fluid
from the second-order integral equations theory}

\author[O.~Pizio, Z.~Soko\l owska, S.~Soko\l owski]{O.~Pizio\refaddr{label1},
Z.~Soko\l owska\refaddr{label2},
S.~Soko\l owski\refaddr{label3}}

\addresses{\addr{label1}
Instituto de Qu\'{i}mica de la UNAM, Coyoac\'{a}n 04510, M\'{e}xico
\addr{label2}
Department of Physical Chemistry of Porous Materials,
 Institute of Agrophysics Polish Academy of Sciences,
Do\'swiadczalna 4,  20--290 Lublin, Poland
\addr{label3}
Department for the Modeling of Physico-Chemical Processes,
Maria Curie-Sk{\l}odowska University,
20--031 Lublin, Poland}

\date{Received April 18, 2010, in final form July 27, 2010}
\begin{document}
\maketitle

\begin{abstract}
 We have studied the  structure and thermodynamic properties of isotropic
three-dimensional core-softened  fluid by using
the second-order Ornstein-Zernike integral equations completed
by the hypernetted chain and Percus-Yevick closures.
The radial distribution functions are compared with those
from singlet
integral equations and with computer si\-mu\-la\-tion data.
 The limits of the region of density anomaly resulting from
 different approximate theories are established. The obtained
 results show that the second-order hypernetted chain approximation
 can be used to describe both the structure
 and the density anomaly of this model fluid. Moreover, we present the
 results of calculations of the bridge functions.


\pacs 61.20.Ja, 61.25.-f, 64.70.pm, 65.20.Jk

\end{abstract}

\section{Introduction}

Recently, significant research activity has been focused on the so-called
core-softened  models in which
the repulsive part of the intermolecular interaction potential
 exhibits a softening region at ``intermediate'' interparticle
separations, in addition to usual hard or soft
 (as in the case of Lennard-Jones (12,6) potential) repulsive branch
at short distances.
 The softening region can be described by a
linear or nonlinear ramp, a shoulder, a single or multiple
attractive well, or a combination of all these
features~\cite{hem,stil,cho1,jag1,sadr,egor,male1,kum,chak,pies,chaim}.
Model fluids with spherically symmetric  core-softened potentials
exhibit anomalous thermodynamic and dynamic behaviours
\cite{nigel1,nigel2,yan}
 that usually occur in real fluids
with directional interparticle interactions, e.g.
water, silica, some liquid metals and phosphorus.
 Examples of the anomalous behavior include the existence
of a density maximum as a function of temperature, an
increase of the diffusion coefficient upon compression,
and, for some systems, the
 existence of multiple fluid-fluid transitions.

 The reason why  the soft-core
model potentials
can  reproduce the density anomaly  is that
at  low densities and temperatures, the neighboring
 molecules are separated by
  distances coinciding  with the range of
 attractive potential well to minimize the
energy. At higher temperatures, the particles can penetrate into
the energetically less favorable softened core to gain more
entropy, thus giving rise to an anomalous contraction upon
heating, see e.g.,~\cite{mittal,netz}. The systems with
core-softened
 potential have been studied using computer simulations
\cite{jag1,sadr,egor,male1,kum,chak,pies,chaim,nigel1,nigel2,yan,mittal,netz,grib,lomba,franio1,male2,evy}
 as well as integral equations methods~\cite{egor,male1,kum,chak,mittal,lomba,franio1,male2}.

Among different models of the core-softened potentials, the
potential used by Barros de Oliveira et
al.~\cite{barro1,barro2,barro3} that yields both the liquid-gas
phase transition and thermodynamic, structural and dynamic
anomalies seems to be interesting. The  potential is the sum of a
Lennard-Jones potential of the well depth
$\varepsilon_{\mathrm{f}}$
 and a Gaussian barrier
centered on radius $r = r_0$ with
 the height of
 $a\varepsilon_{\mathrm{f}}$,
\begin{equation}
u(r)=4\varepsilon_{\mathrm{f}}\left[ \left( \frac{\sigma_{\mathrm{f}}}{r}\right)^{12} -
\left( \frac{\sigma_{\mathrm{f}}}{r}\right)^{6} \right] + a\varepsilon_{\mathrm{f}}
\exp\left[ -\frac{1}{c^2}\left( \frac{r-r_0}{\sigma_{\mathrm{f}}}\right)^2\right].
\end{equation}
Depending on
the choice of  parameters, equation~(1)
  represents the whole family of two
length scales intermolecular interactions, from a deep double-well
potential~\cite{cho1,cho2,netz2} to a repulsive
shoulder~\cite{jag1,jag2,jag3}. Moreover, due to its continuity,
the potential (1)
 is convenient both for the
 Monte Carlo and molecular dynamics simulations.

For specific choices of the parameters, the potential (1) can
exhibit a double well behavior, similar to the potential studied
by Cho et al.~\cite{cho1,cho2}. However, the attractive double
well may bring both the liquid-gas phase transition and the
anomalies to higher temperatures into an unstable region of the
phase diagram in the pressure-temperature plane~\cite{netz2}. In
order to circumvent this behavior, several studies
\cite{barro1,barro2,barro3,ps1,ps2,ps3,charusita} were carried out
for a specific set of  parameters that give rise to a potential
with a repulsive ramp followed by a small attractive well. In
particular, the following values were often used: $a=5$, $c=1$ and
$r_0=0.7\sigma_{\mathrm{f}}\,$. It was shown
\cite{barro1,barro2,barro3} that for the above set of parameters,
the pressure-temperature curves have a minimum if calculated at
certain densities (i.e. $(\partial P/\partial T)_{\rho}=0$).
Consequently, the derivative $(\partial \rho/\partial
T)_{\mathrm{P}}=0$ for some thermodynamic states. The region of
density anomaly corresponds to state points for which $(\partial
\rho/\partial T)_{\mathrm{P}} > 0$ and is bounded by the locus of
points for which the thermal expansion coefficient is equal to
zero.

In our recent work~\cite{ps3} we applied grand canonical Monte
Carlo simulation and integral equations with hypernetted chain
(HNC), as well as Rogers-Young closures to study the above defined
model. We found still another anomaly in the system, namely we
demonstrated that for some temperatures the derivative of the
density with respect to the chemical potential exhibits a minimum
at a certain density, followed by a maximum. Also, a peculiarity
in the dependence of the specific heat upon density was found. We
compared the pair distribution functions resulting from integral
equations and from computer simulations and established that
common HNC approximation was not successful in capturing the
region of anomalies in contrast to Rogers-Young approximation that
imposes thermodynamic self-consistency by construction. However,
the HNC was very accurate at high fluid densities. It is worth
mentioning that a popular mean spherical approximation has not
been used to study the model in question because this approach
intrinsically requires an adequate splitting of the potential into
a short- and long-ranged parts. This approximation does not seem
to be beneficial for some soft-core models with two length scales
and of this study in particular.

The results of integral equations reported in the literature for
the potential (1) were based on the singlet OZ equation~\cite{ha}.
A more sophisticated approach, but one that is also more demanding
of computational resources, is based on the inhomogeneous OZ
equation~\cite{hen}.

The inhomogeneous OZ equation has usually been used not only to
calculate the structure and thermodynamic properties of fluids in
contact with surfaces~\cite{sokoOZ,sokoOZ1,hendeOZ,troOZ,pizioOZ},
but also to determine the structure and the surface tension at the
liquid-vapor interface~\cite{kovalOZ}. However, one can also
assume that for a single-component fluid, the source of an
external potential field is just a single distinguished particle
identical to all remaining molecules of the system. This is the
so-called Percus' trick~\cite{percus1,percus2} or the ``test
particle'' method. The method  was extended later by
Attard~\cite{attard1,attard2} in the framework of the second-order OZ
approach for bulk fluids, see e.g.~\cite{dough1,myrosio1,myrosio2}
for the discussion of related important issues. Moreover, this
approach was successfully applied to several model fluids with
spherically symmetric intermolecular
potentials~\cite{dh1,dh2,dh3,dh4,dh5,dh6,bry}. Better description of the
desired properties or in some cases even qualitatively new
findings in comparison to the singlet level theory were obtained.
However, to our best knowledge, this approach has never been
tested for the core-softened models of the interparticle
potentials.
Therefore, the motivation of this work is to use the second-order
integral equations to the fluid of particles interacting via the
core-softened potential (1). The second-order equations are
applied to the study of the microscopic structure and
thermodynamic properties of the fluid, in particular, in view of
its anomalous behavior in a certain region of thermodynamic
states. The results are compared with computer simulation data and
with the predictions of the singlet integral equations~\cite{ps3}.
We explore here the second-order Percus-Yevick and  hypernetted
closures to the nonuniform OZ equation.

\section{Theory}
The OZ equation for an inhomogeneous fluid, wherein the
density $\rho(\mathbf{r})$ is not constant reads
\begin{equation}
\label{eq:OZ}
h_2(\mathbf{r}_1,\mathbf{r}_2)=c_2(\mathbf{r}_1,\mathbf{r}_2)
+\int \rd\mathbf{r}_3 h_2(\mathbf{r}_1,\mathbf{r}_3)c_2(\mathbf{r}_3,\mathbf{r}_2)
 \rho(\mathbf{r}_3),
\end{equation}
where $h_2(\mathbf{r}_1,\mathbf{r}_2)$ and $c_2(\mathbf{r}_1,\mathbf{r}_2)$
are the total and direct correlation functions of an nonuniform fluid.
We have used here the subscript ``2'' in order to
distinguish these functions from the common uniform fluid correlation functions.

Generally, equation~(\ref{eq:OZ}) applies when
the inhomogeneity is due to  an external potential field.
However, one distinguished  molecule
can be also considered as a source of the external potential.
 In such cases equation~(\ref{eq:OZ}) can be solved using any of the
common approximations, as HNC
 \begin{equation}
 \label{eq:HNC2}
 h_2(\mathbf{r}_1,\mathbf{r}_2)=\exp[h_2(\mathbf{r}_1,\mathbf{r}_2)
 -c_2(\mathbf{r}_1,\mathbf{r}_2)-\beta u(|\mathbf{r}_1-\mathbf{r}_2|)
 ]-1,
 \end{equation}
 or Percus-Yevick (PY) approximation
 \begin{equation}
 \label{eq:PY2}
 c_2(\mathbf{r}_1,\mathbf{r}_2)=\{1-\exp[\beta u(|\mathbf{r}_1-\mathbf{r}_2|)]\}
 [h_2(\mathbf{r}_1,\mathbf{r}_2)+1].
 \end{equation}
In the above  $u(|\mathbf{r}_1-\mathbf{r}_2|)$ is the pair potential and $\beta=1/kT$
is the inverse temperature. However, we stress that the above closures do not relate
 uniform, but rather nonuniform correlation functions and therefore
we refer to the results of equation~(\ref{eq:OZ}) with the closure
given by equation~(\ref{eq:HNC2}), or by equation~(\ref{eq:PY2}),
as to the HNC2 and PY2 results, respectively, in contrast to the
usual HNC and PY theories~\cite{ha}.

To have a set of equations completed we also need a relation
between $\rho(\mathbf{r})$ and the pair correlation functions. The
exact equation developed by Lovett, Mou, Buff and Wertheim
(LMBW)~\cite{LMBW,WE} reads
\begin{equation}
\label{eq:LMBW}
\nabla \ln[y(\mathbf{r}_1)] = \int \rd\mathbf{r}_2 c_2(\mathbf{r}_1,\mathbf{r}_2)
\nabla \rho(\mathbf{r}_2),
\end{equation}
where $y(\mathbf{r}_1)$ is the one-particle background (cavity) function,
 $\rho(\mathbf{r}_1) = \exp[-\beta v(\mathbf{r}_1)]y(\mathbf{r}_1)$ and $v(\mathbf{r}_1)$
 is the potential due to the particle which is regarded as the source of the
 inhomogeneity. In our case, $v(\mathbf{r}_1)$ is just the pair potential,
 $v(\mathbf{r}_1) \equiv u(|\mathbf{r}_1-\mathbf{0}|)$, where the distinguished
 particle is set at the origin, $\mathbf{r}_2=\mathbf{0}$.

One should mention here an important difference between the
theories used to study systems in contact with a
surface~\cite{sokoOZ,sokoOZ1,hendeOZ,troOZ,pizioOZ} (or the
systems involving a gas-liquid interface~\cite{kovalOZ}) and
theories based on the method of Attard. In the former case the
two-particle correlation functions reduce to the functions for
bulk fluids, providing the both particles to be
 located far away from an inhomogeneous region.
In the Attard's approach, the function $h_2$  corresponds rather to the third-order
(conditional)
correlation function of a bulk fluid, while the ``usual'' pair correlation function
is related to the local density.
Indeed, because the bulk pair distribution function, $g(r)$,
 gives the probability density of
finding a pair of fluid molecules at a separation, $r$, the
$g(r)$ is
related to the local density, $\rho(\mathbf{r})$, via
\begin{equation}
\label{eq:local}
g(r)=\rho(|\mathbf{r}|)/\rho,
\end{equation}
where $|\mathbf{r}|$ is the distance from the distinguished particles that is
the source of the system inhomogeneity and $\rho=\lim_{|\mathbf{r}|\to \infty}
\rho(|\mathbf{r}|)$ is the bulk density. Of course, the cavity function $y(r)$ that enters
equation~(\ref{eq:LMBW}) satisfies the relation $g(r)=\exp[-\beta u(r)]y(r)$.

To solve equations~(\ref{eq:OZ}) and (\ref{eq:LMBW}) with the closure given by
 equation~(\ref{eq:HNC2}) or equation~(\ref{eq:PY2})
we use the numerical algorithm Attard's~\cite{attard1} that relies
on the expansion of the two-particle functions ($h_2$ and $c_2$)
 in series of Legendre polynomials, for details see~\cite{attard1}.

Note that the closure equations (\ref{eq:HNC2}) and (\ref{eq:PY2}) constitute
an approximation, similar to the common bulk theory, whereas the OZ equation
(\ref{eq:OZ}) and the equation for the profile, (\ref{eq:LMBW}) are exact.

Equation~(\ref{eq:LMBW}) satisfies the relation $g(r)=\exp[-\beta u(r)]y(r)$.
 The bulk  direct
correlation function, $c(r)$, can be recovered from $g(r)$ using the bulk Ornstein-Zernike
(OZ) equation
\begin{equation}
\label{eq:bOZ}
h(|r_{12}|)-c(|r_{12}|)=\rho\int \rd\mathbf{r}_3h(|r_{13}|)c(|r_{23}|),
\end{equation}
where $h(r)=g(r)-1$.

The knowledge of the total, $h(r)$, and the cavity, $y(r)$,
functions enable us to determines the bridge function, $B(r)$. The
bridge function is defined as, see e.g.~\cite{ha,yukh}
\begin{equation}
B(r)= \ln y(r) - \gamma(r),
\end{equation}
where $\gamma(r)=h(r)- c(r)$. The bridge function plays a key role
in theories based on the singlet OZ equation (7). It is  given by
the smallest set of diagrams in the graphical expansion, and all
other functions can be calculated from it via the OZ
equation~\cite{ha}.

The main difference between the first- and the second-order
integral equation theories for bulk fluids is that in the latter approach,
the bridge function results from the calculations performed, whereas
in the case of first-order theories the bridge function is imposed
as a closure.
Previous calculations carried out
for hard spheres, as well as for Lennard-Jones fluids, gave
insight into the course of the bridge functions. However, to
our best knowledge no results of calculations of the
bridge function for core-softened potential
models have been presented so far.

\section{Results and discussion}

The second-order integral equations were solved using the Attard's
algorithm ~\cite{attard1}. Usually, 80 Legendre polynomials were
used and the grid in $r$ was $0.04\sigma$. To test the accuracy we
increased the number of Legendre polynomials to 120 and decreased
the grid size to $0.025\sigma$. Moreover, we also solved the
singlet integral equations and carried out grand canonical
ensemble Monte Carlo simulations. The details on the two latter
methods are presented in our previous work~\cite{ps3}. We also
note that the pressures given below were calculated from the
virial equation of state (see equation~(6) of~\cite{barro2}).

\begin{figure}[ht]
\begin{center}
\includegraphics[width=0.4\textwidth]{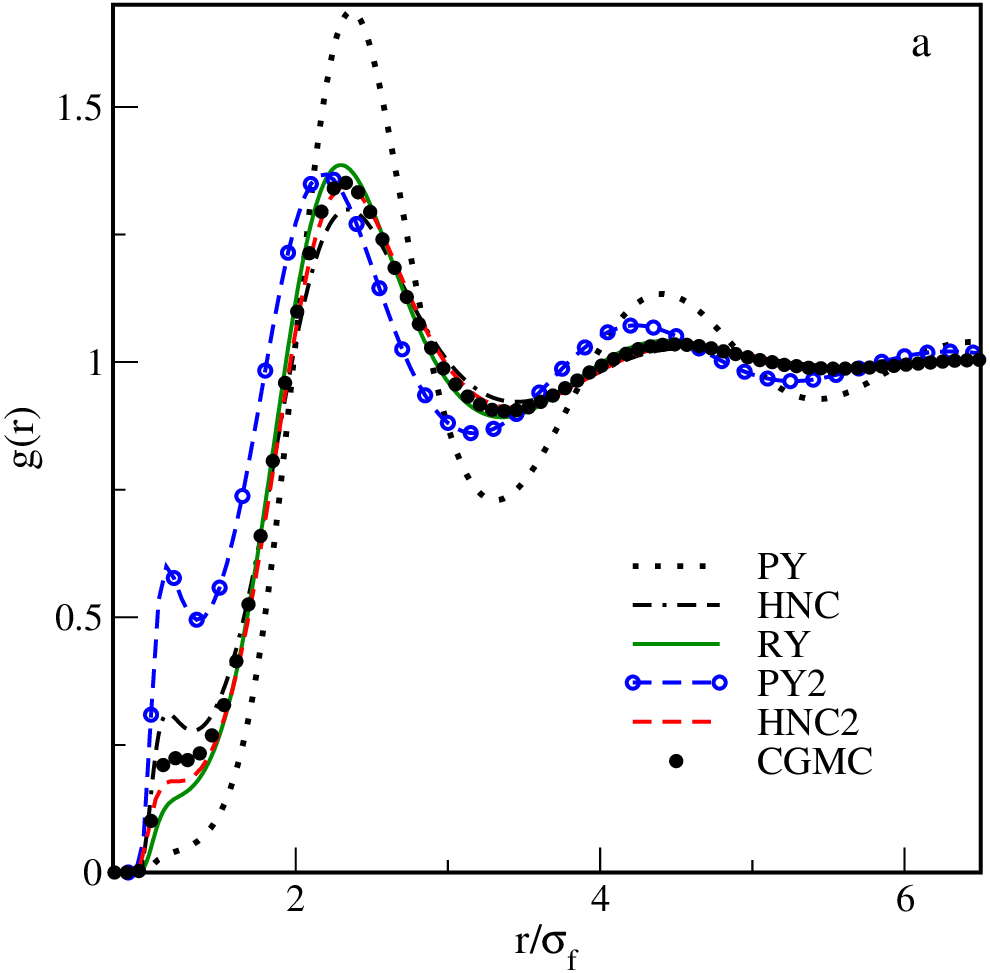}
\includegraphics[width=0.4\textwidth]{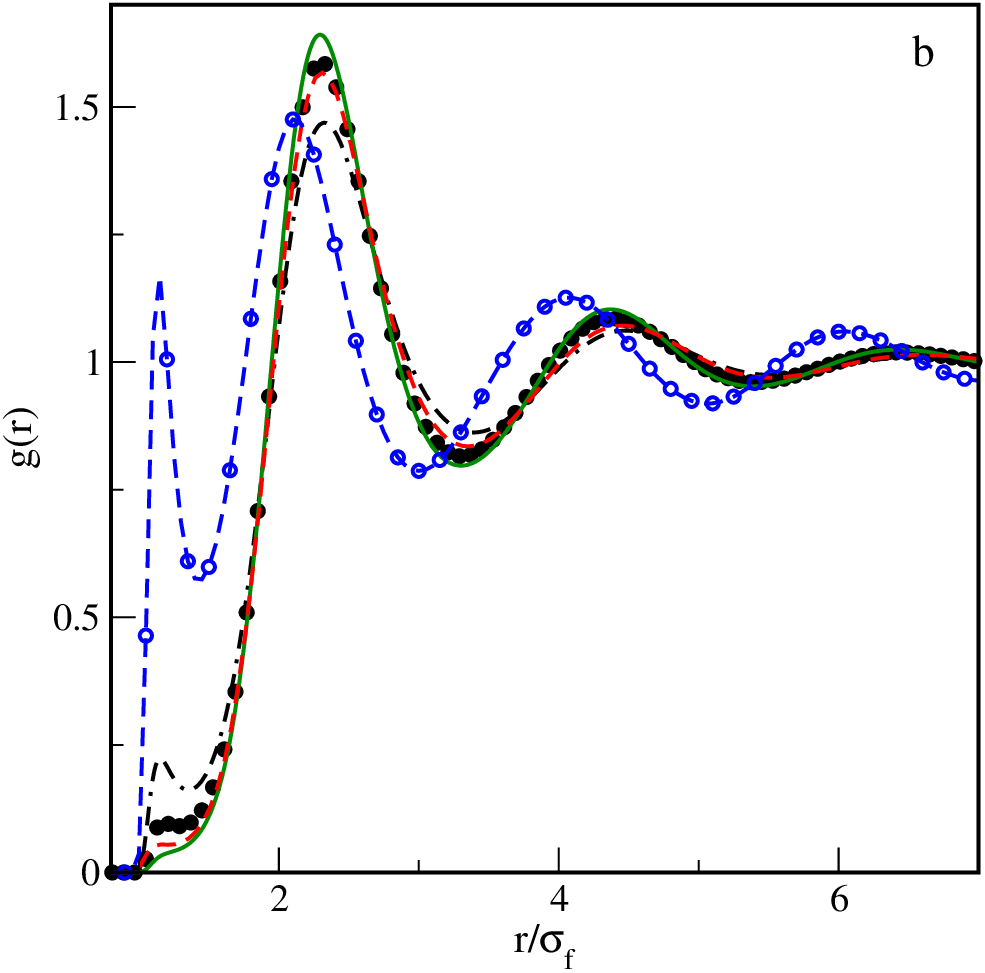}
\end{center}
\begin{center}
\includegraphics[width=0.4\textwidth]{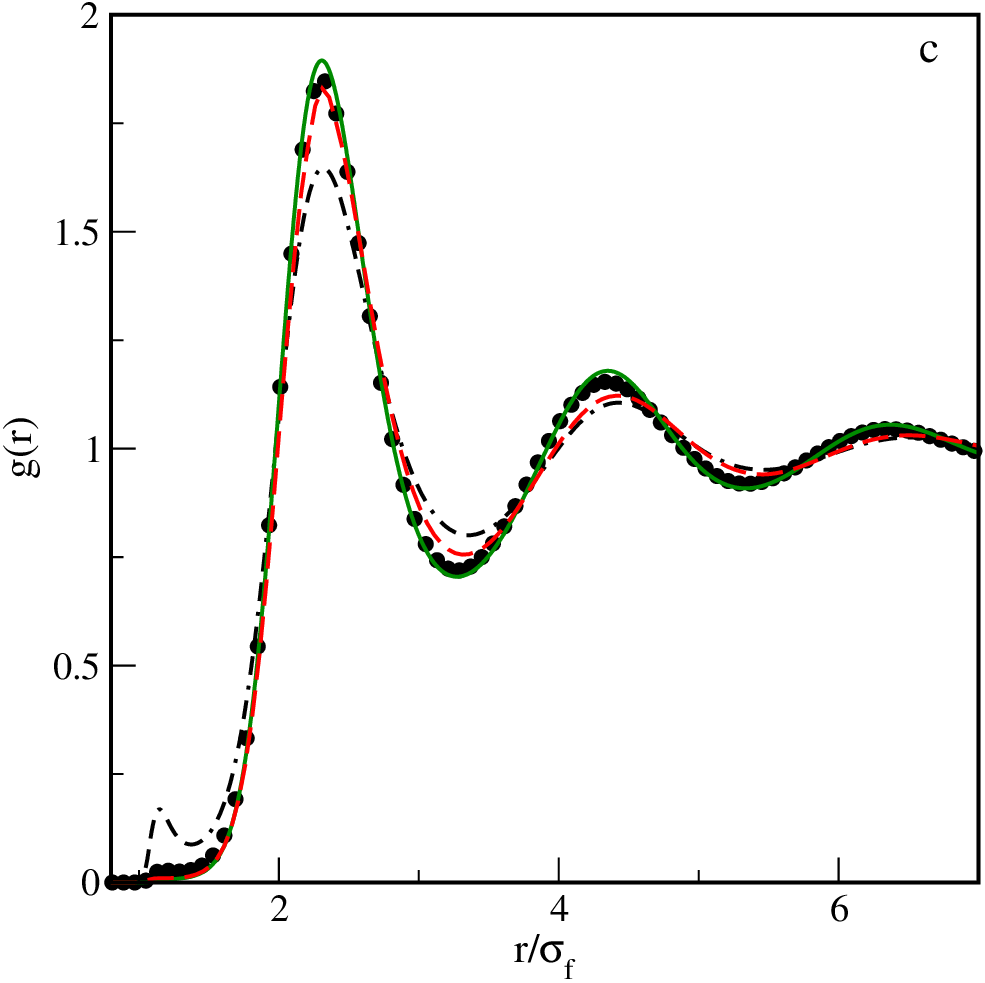}
\end{center}
\caption{
(Color on-line) Radial distribution functions evaluated from
the singlet (PY, HNC and RY), second-order (PY2, HNC2) theories and from GCMC
simulation at $\rho^*=0.1$ and at $T^*=0.5$ (part~a), 0.3 (part~b) and 0.2 (part~c).
 The nomenclature of the lines is given in part~a. Note that consecutive
 parts b and c contain the results of selected theories.}
\end{figure}
Similarly to our previous work\cite{ps3}, all the calculations
have been carried out assuming that the parameters of the
potential (1) are $a=5$, $c=1$ and $r_0=0.7\sigma_{\mathrm{f}}\,$.
First, we concentrate on a comparison of the pair distribution
functions obtained from the second-order OZ equation, the singlet
OZ equation and the Grand Canonical ensemble (GCMC) computer
simulations~\cite{ps3}. Comparisons are for thermodynamic
conditions that cover
 the density anomaly region found in molecular dynamics studies~\cite{barro2}.
In figures~1~(a)--(c) we display the functions $g(r)$ evaluated at
the density $\rho^*=\rho\sigma_{\mathrm{f}}^3=0.1$ and at three
temperatures, $T^*=kT/\varepsilon_{\mathrm{f}}=0.5$, 0.3 and 0.2.
At the highest considered temperature the results of the singlet
theories (PY, HNC and Rogers-Young, (RY)), the second-order
theories, PY2 and HNC2 are compared with GCMC data. Before
discussing the results shown in figure~1, we recall that  the
singlet RY closure contains one adjustable parameter and this
parameter is evaluated requiring the equality of pressures from
the virial and compressibility routes, for details see~\cite{ps3}.

In figure~1~(a) we see that the singlet PY approximation yields the
radial distribution function that
significantly deviates from all remaining results.
The second-order PY2 approximation is definitely more
accurate than the singlet PY theory, but its
results still differ much from computer simulations.
In contrast to the singlet PY theory, the singlet HNC closure
 yields the results that qualitatively agree with computer simulations,
 but among all the singlet theories the thermodynamically self-consistent
 RY closure gives the best predictions. For the thermodynamic
 state from figure~1~(a)
the second-order  hypernetted chain approximation reproduces
the GCMC data with the highest precision. We stress that in contrast to the
singlet RY theory the HNC2 approach
contains no adjustable
parameter.

\begin{figure}[!h]
\begin{center}
\includegraphics[width=0.4\textwidth]{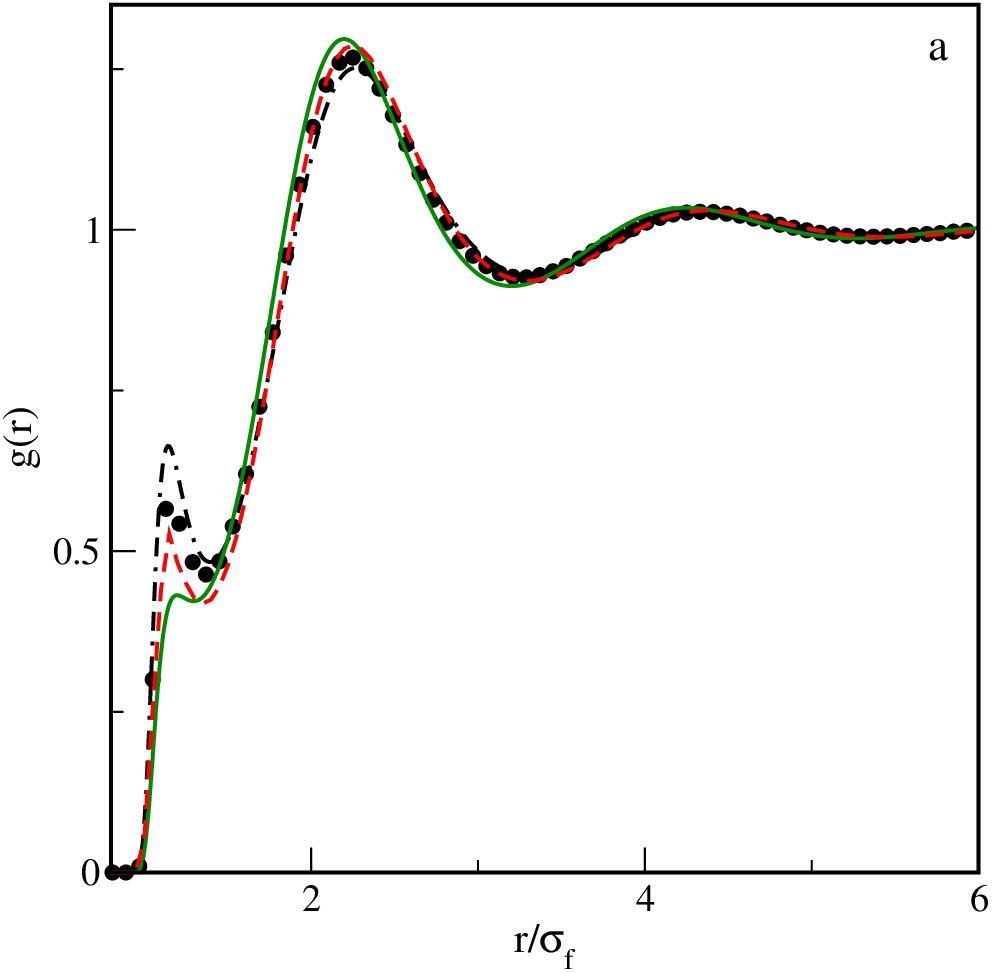}
\includegraphics[width=0.4\textwidth]{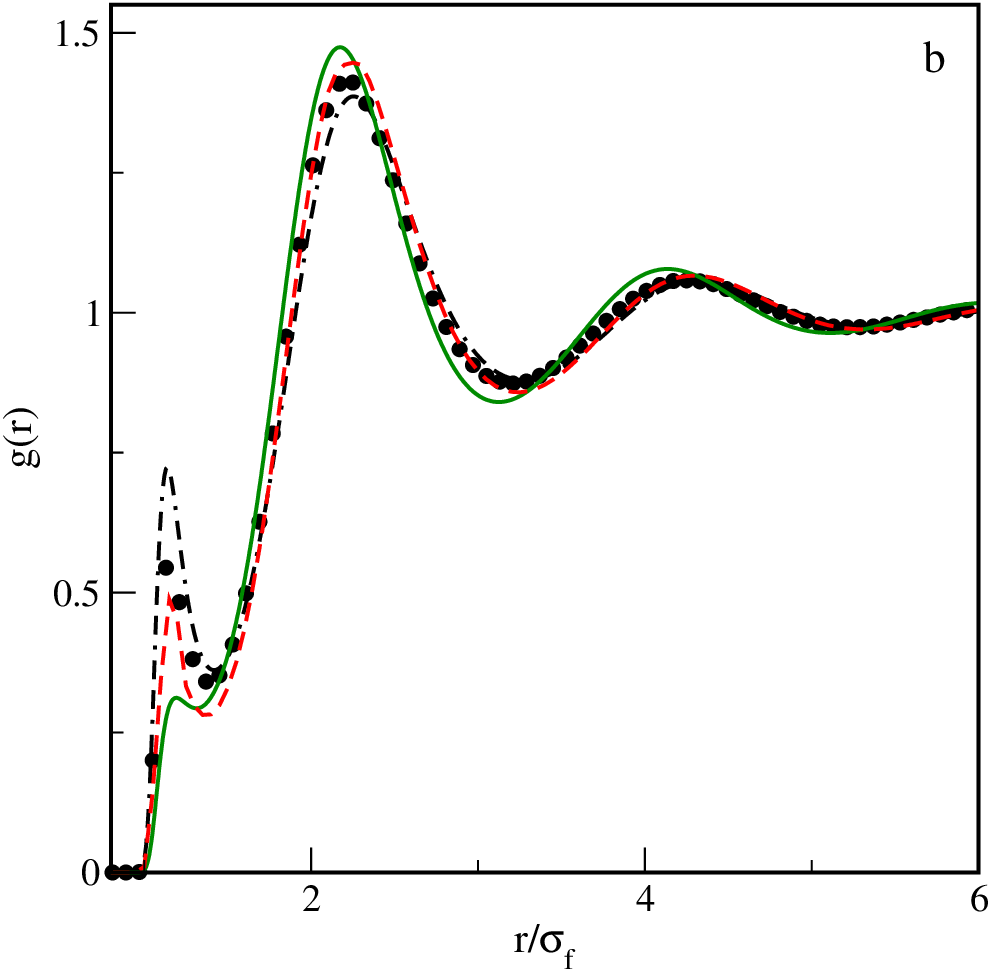}
\end{center}
\begin{center}
\includegraphics[width=0.4\textwidth]{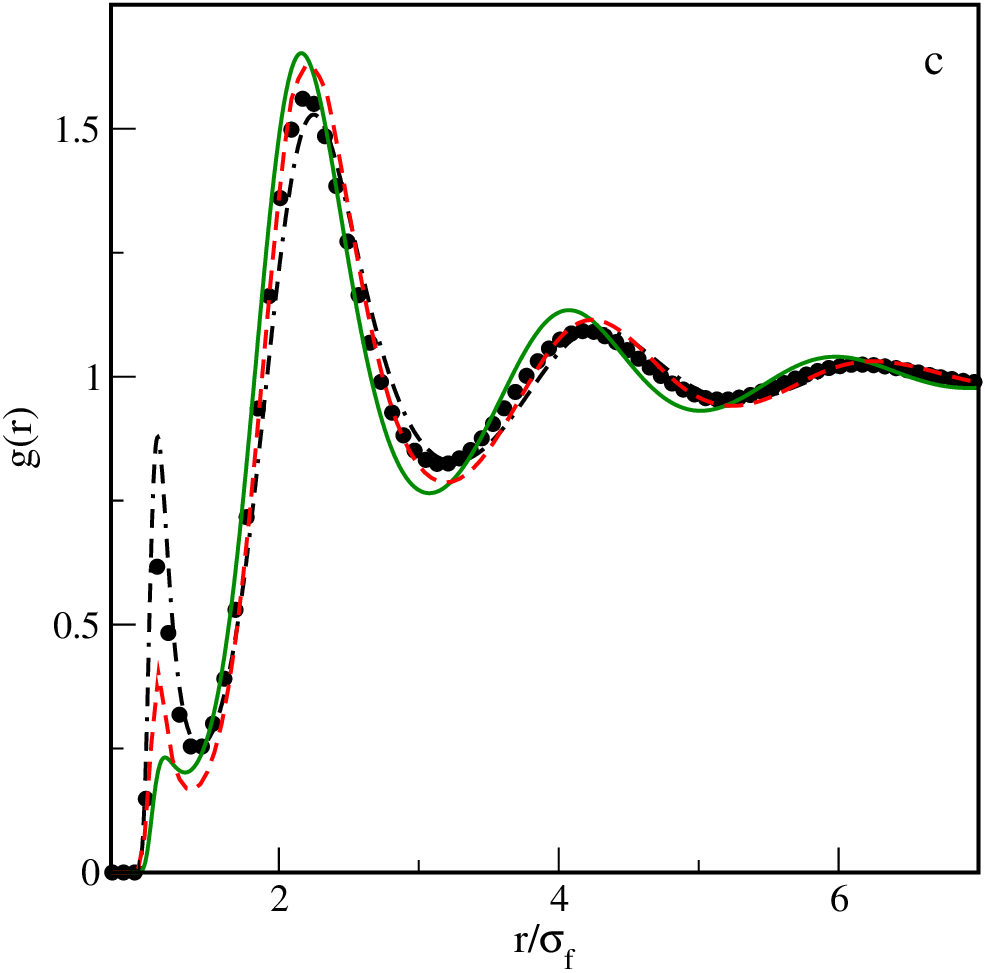}
\includegraphics[width=0.4\textwidth]{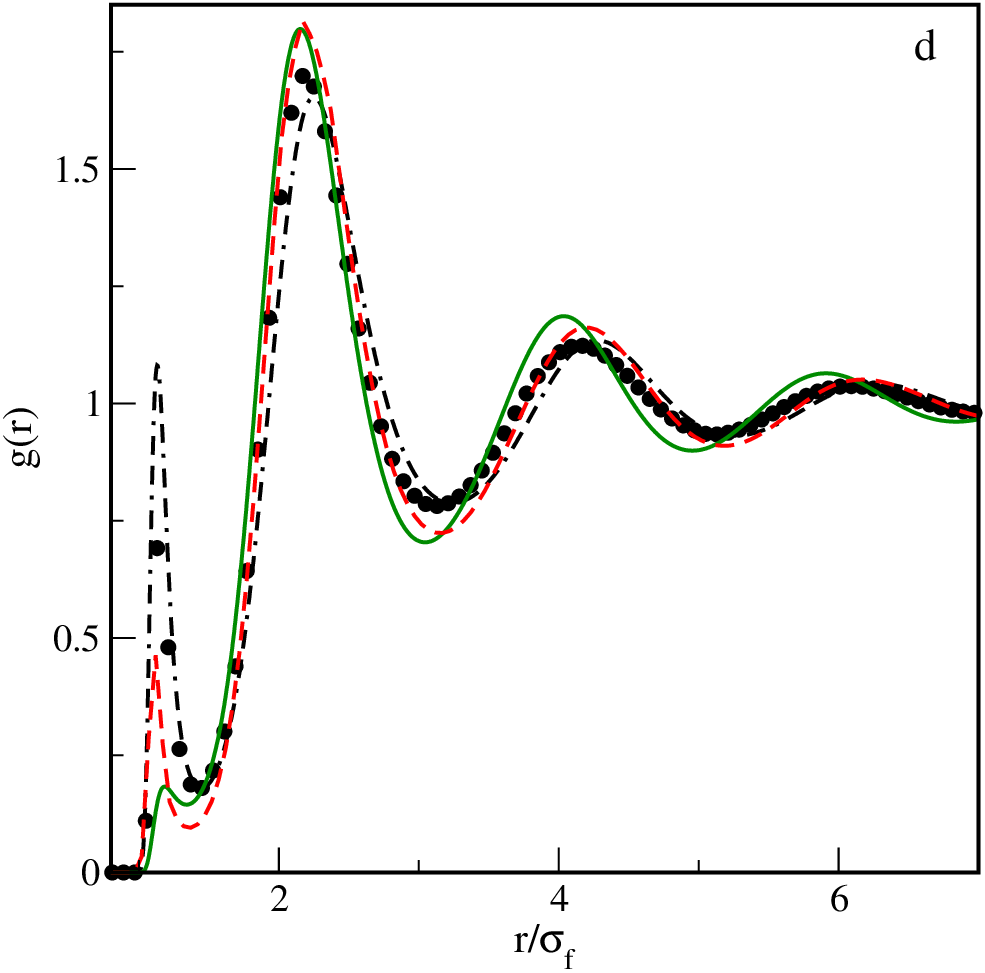}
\end{center}
\caption{
(Color on-line) Radial distribution functions evaluated from
the singlet (HNC and RY), second -order (HNC2) theories and from GCMC
simulation at $\rho^*=0.14$ and at $T^*=0.5$ (part~a), 0.3 (part~b), 0.2 (part~c)
and 0.15 (part~d).
 The nomenclature of the lines is given in figure~1~(a).}
\end{figure}
\begin{figure}[!h]
\begin{center}
\includegraphics[width=0.4\textwidth]{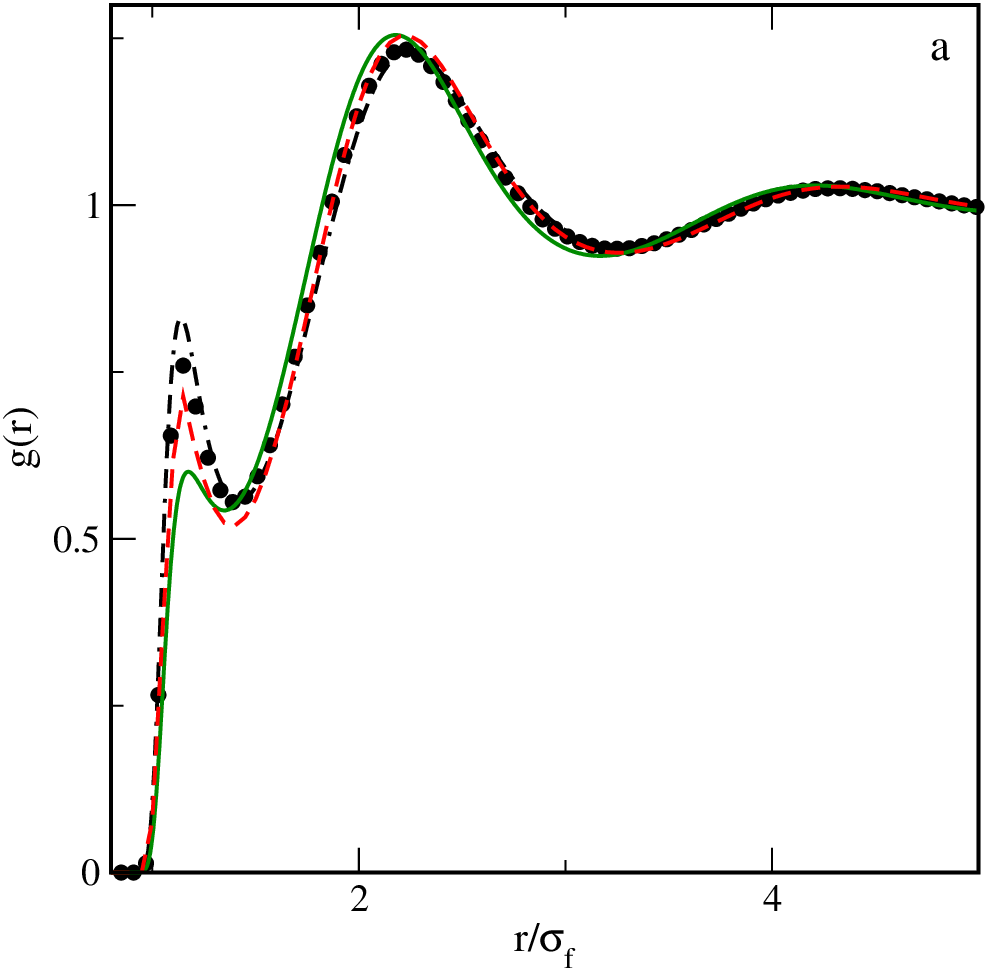}
\includegraphics[width=0.4\textwidth]{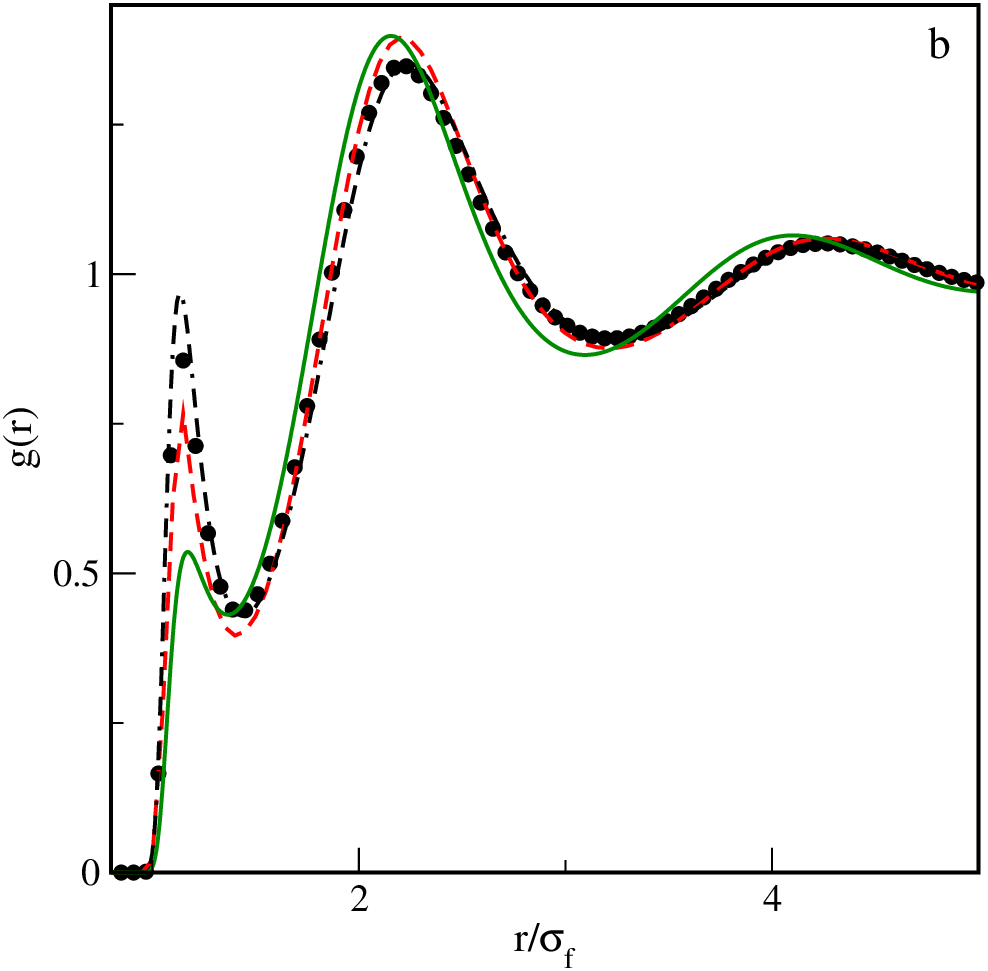}
\end{center}
\begin{center}
\includegraphics[width=0.4\textwidth]{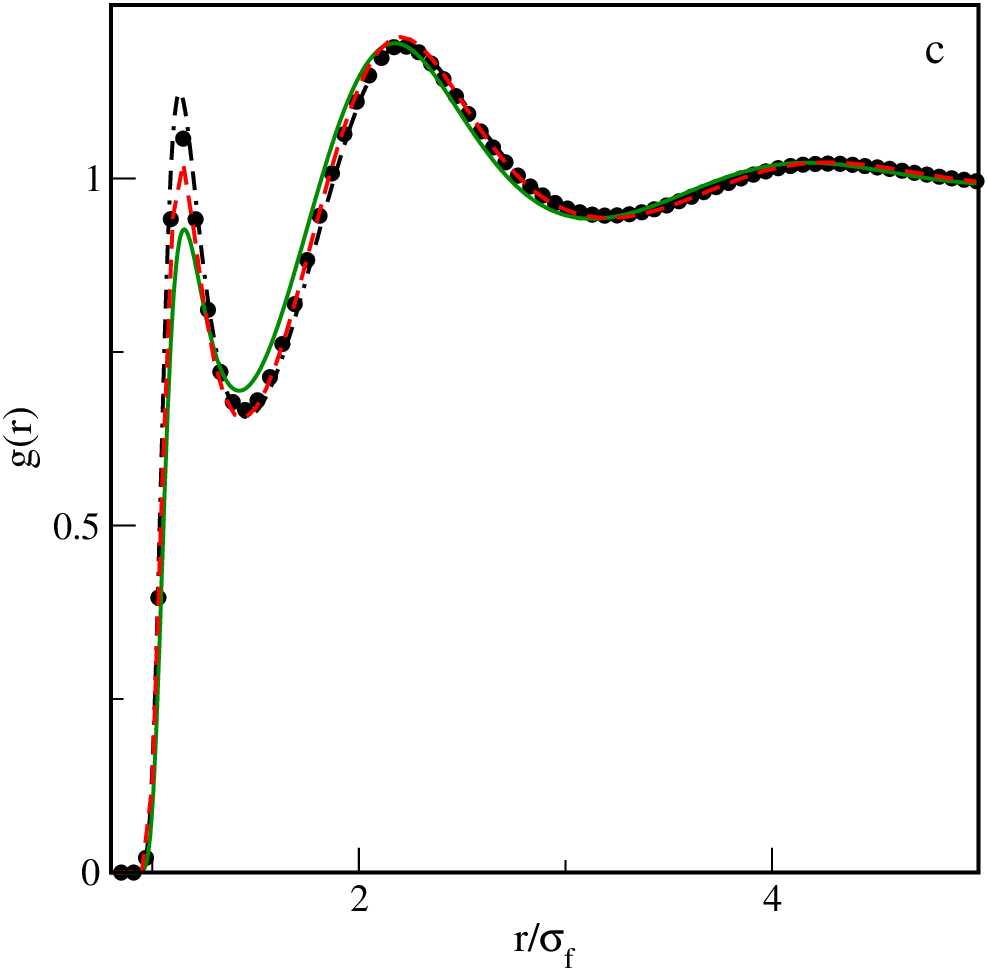}
\includegraphics[width=0.4\textwidth]{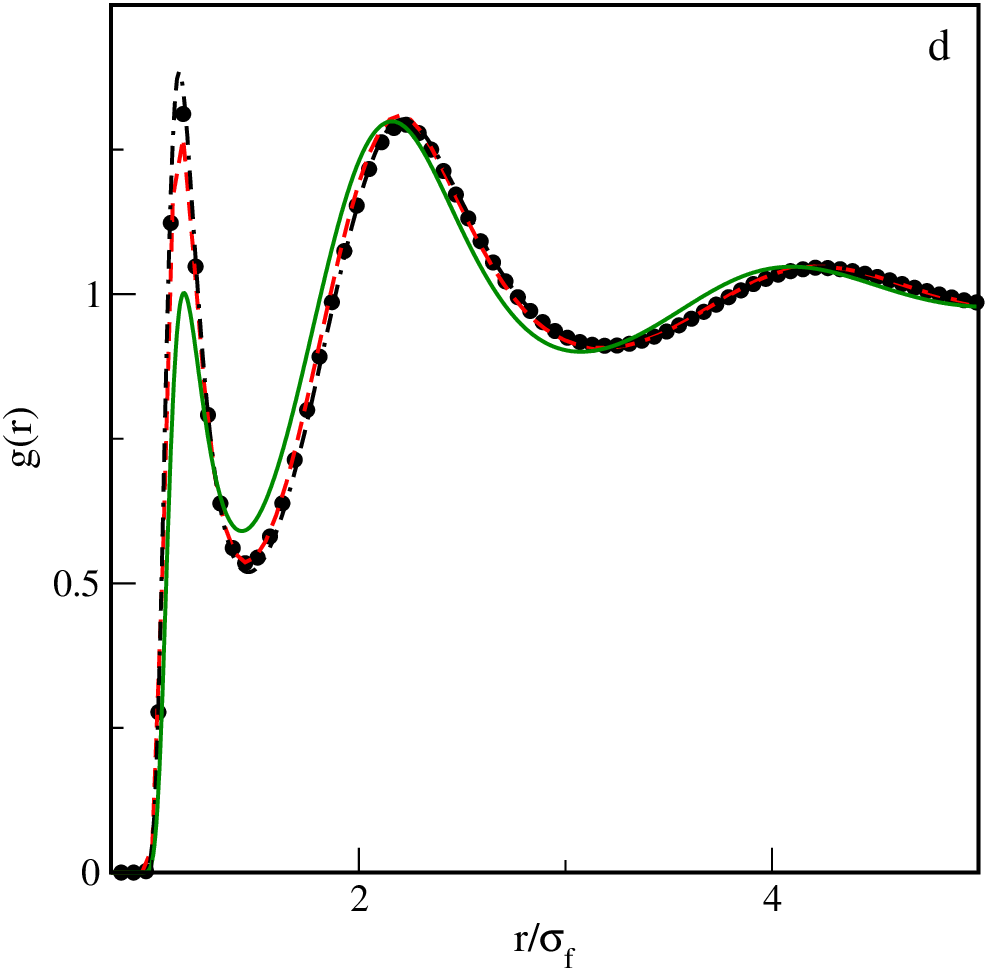}
\end{center}
\caption{
(Color on-line) Radial distribution functions evaluated from
the singlet (HNC and RY), second -order (HNC2) theories and from GCMC
simulation at $\rho^*=0.16$ and at $T^*=0.5$ (part~a), $\rho^*=0.16$ and at $T^*=0.3$ (part~b),
 $\rho^*=0.2$ and at $T^*=0.5$ (part~c), and  $\rho^*=0.2$ and at $T^*=0.3$ (part~d).
 The nomenclature of the lines is given in figure~1~(a).}
\end{figure}
The accuracy of the second-order HNC2 theory is also
confirmed by the results presented  in figure~1~(b).
HNC2 theory well reproduces the location and the height of the first maximum
of $g(r)$ and the
shoulder at distances $r/\sigma_{\mathrm{f}} < 1.4$. However, the first
minimum of $g(r)$ at $r/\sigma_{\mathrm{f}}\approx 3.32$ is a bit better
predicted by the RY
approximation. Similar conclusions
can be drawn from the inspection of the curves presented in figure~1~(c).
Since the PY2 approximation actually fails to reproduce the simulation data with a reasonable accuracy,
 the PY and PY2 results are omitted in figure~1~(c) and in the following figures~2 and 3.
The singlet PY approximation is quite accurate when applied to a
hard-sphere system~\cite{ha}. On the contrary, in the case of soft
potentials, e.g., Gaussian-like potential, the singlet HNC
approximation works quite satisfactory~\cite{gauss}. It is thus
not surprising that for core-softened potential, the hypernetted
chain approximation (at singlet and second-order levels) performs
better than the the Percus-Yevick closures. However, the observed
 big differences
between the predictions of these approximations are surprising,
because in the case of the systems studied so
far~\cite{dh1,dh2,dh3,dh4,dh5,dh6} the differences between the PY2
and HNC2 approximations were rather small~\cite{henderson}.

Further tests of the accuracy of the HNC2 approximation are shown
in figures~2 and 3. Figure~2 shows the results at the density
$\rho^*=0.14$, consecutive parts {\it a}~--~{\it d} are for the
temperatures $T^*=0.5$, 0.3, 0.2 and 0.15, respectively. Again,
the HNC2 approximation is superior over the best singlet RY
theory. In particular, it better describes the formation of the
first peak of $g(r)$ within the  repulsive ramp of the potential
(1) (the plot of the potential (1) gives figure~1
of~\cite{barro2}), though both RY and HNC2 theories underestimate
the height of this maximum. By contrast, the singlet HNC approach
describes much higher first peak of $g(r)$ compared to GCMC data.

Finally, we calculated the radial distribution functions at higher densities,
$\rho^*=0.16$ and 0.20 (figure~3). For these densities, the second-order HNC2 equation
performs well.
In particular, its predictions of the height of consecutive
maxima and minima of $g(r)$ are more accurate than the singlet RY approximation.
However, for some  state points, the accuracy of the
singlet HNC equation is quite good and even better than
the HNC2 theory, cf. figures~3~(b) and 3~(c).

Before discussing an anomalous behavior of the system in the pressure-temperature
plane, that results from the
second-order integral equations, we  briefly recall previous findings.
The line of temperatures of maximum density (TMD) was determined
from NPT molecular dynamics~\cite{barro2} and singlet PY
\cite{barro2} and RY~\cite{barro2,ps3} integral equations. It was
also demonstrated~\cite{barro2,ps3} that the singlet HNC approach
fails to predict this anomaly.

 Molecular
dynamic simulations revealed that for
densities $0.12<\rho^*<0.14$,
the pressure-temperature curves at constant density have a minimum
(see also our comment
in the introductory section) which implies density anomaly.
The maximum temperature at which the density anomaly still exists
is $T^*_{\rm max}\approx 0.25$~\cite{barro2}. The predictions of
the singlet RY approximation  depend on the method according to
which its adjustable parameter was determined~\cite{ps3}.
 Barros de Oliveira et al.~\cite{barro2} obtained the adjustable parameter
 of the RY equation
  by checking the consistency between the compressibilities,
  calculated from the virial and compressibility equations of state.
  However, in our work~\cite{ps3} we imposed the so-called global coexistence
  criterion, according to which the pressures from two above mentioned thermodynamic
  routes were compared. Barros de Oliveira et al.~\cite{barro2} found that
  the density anomaly exists for the  densities from the range of $[0.12,0.14]$
  and that $T^*_{\rm max}\approx 0.23$. Our calculations, however, led to a
  wider interval of the densities, $[0.12,0.19]$. We should also note
  that the singlet PY approximation predicts~\cite{barro2}
   much wider region of densities, at which
  the density anomaly occurs, $[0.13,0.3]$. Instantaneously, this region is
  shifted towards much higher temperatures, with $T^*_{\rm max}= 0.86$.

\begin{figure}[ht]
\centerline{\includegraphics[width=0.65\textwidth]{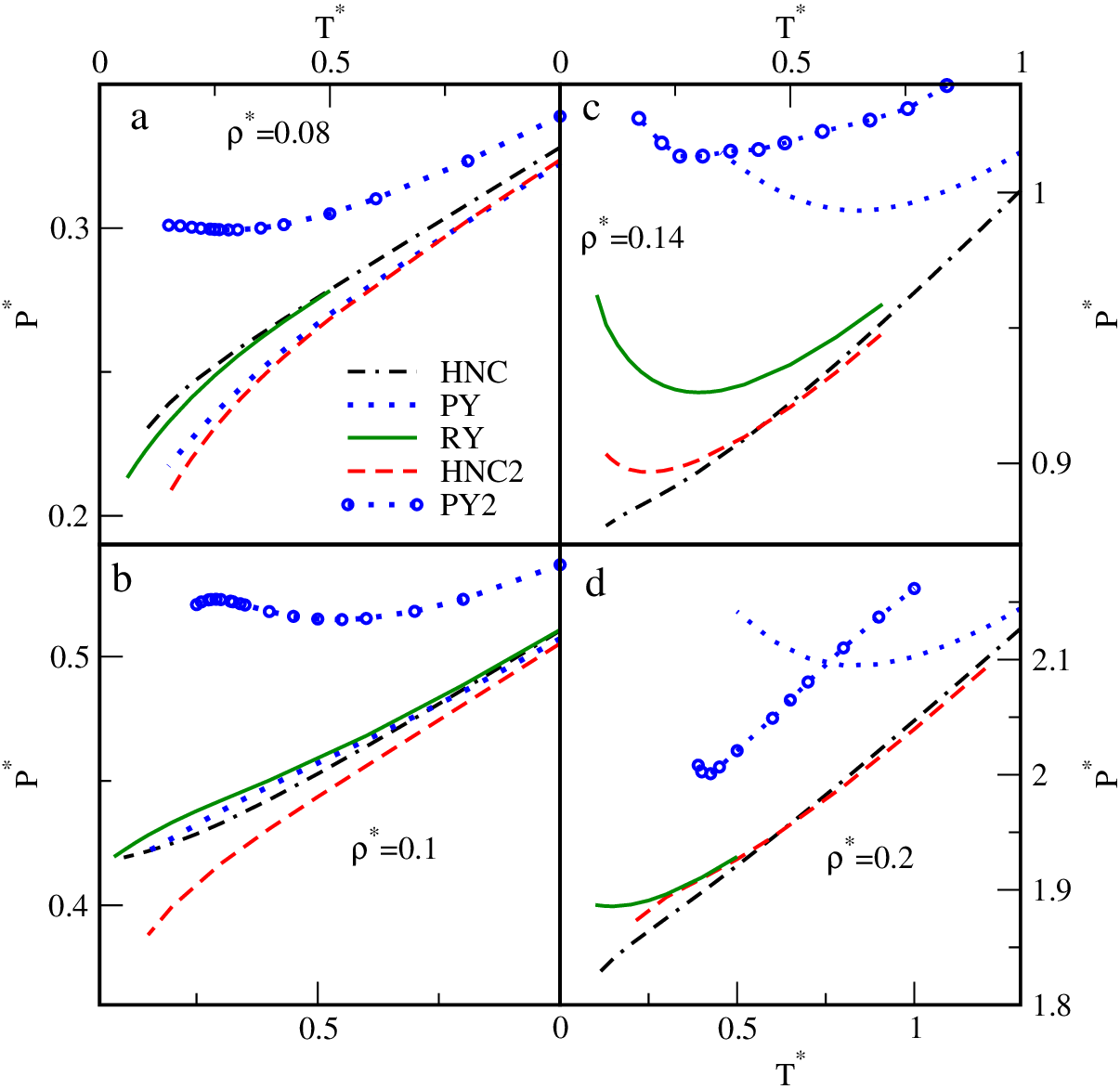}}
\caption{ (Color on-line.) Isochores resulting from different
approximate
 theories  for different densities, given in consecutive parts. The nomenclature of
 the lines is displayed in part~a.}
\end{figure}
\begin{figure}[!h]
\centerline{\includegraphics[width=0.52\textwidth]{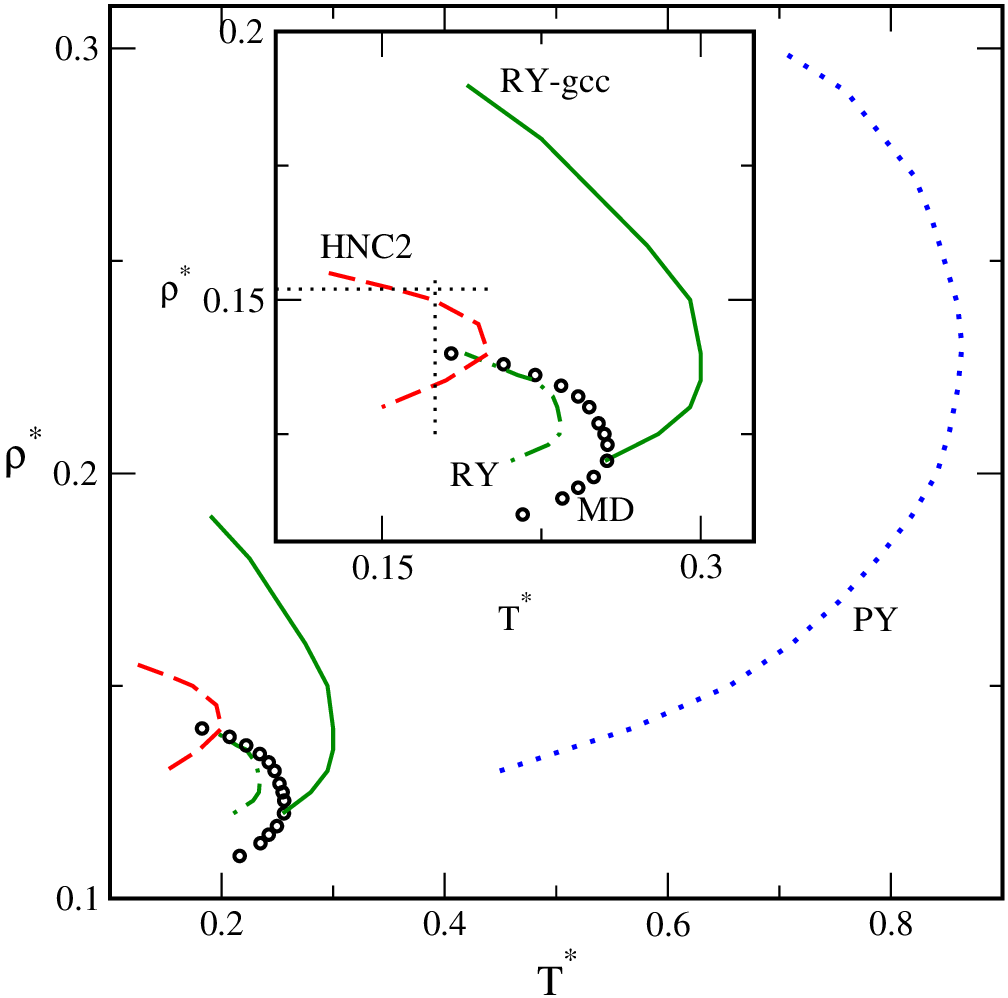}}
\caption{ (Color on-line) The TMD lines from different approximate
theories and from MD simulations~\cite{barro2} (open circles).
 Abbreviations are as follows: HNC2 denotes the second-order hypernetted chain theory results;
 PY is the singlet Percus-Yevick theory~\cite{barro2}, RY denote the Roger-Young results
 obtained by Barros de Oliveira et al.~\cite{barro2} and RY-gcc are the Roger-Young results
 calculated  using the global coexistence criterion. The inset magnifies a part of the
 main figure. Dotted lines here indicate the thermodynamic conditions at which the
 bridge functions displayed in figure~6 were calculated.}
\end{figure}
Figure~4 shows the isochores evaluated from different theories (the reduced pressure is
defined as $P^*=P\sigma_{\mathrm{f}}^3\varepsilon_{\mathrm{f}}$).
At low densities ($\rho^*<0.07$; the relevant curves are not shown for the
sake of brevity)
none of the approximations predicts the existence of the
density anomaly.
For $\rho^*=0.08$ (part~a) only the PY2 theory yields a weak minimum of the pressure,
all the remaining approximations do not show any anomaly. 
However, it is interesting to note that the results of the singlet
PY and HNC2 approximations almost coincide, except for low
temperatures. For $\rho^*=0.1$ (part~b)  only the PY2 isochore
exhibits anomaly, but now the minimum of the pressure is preceded
by a maximum. This behavior is completely unexpected, but, it is
rather an artifact of the theory, because no computer
simulations~\cite{barro2}
 provide its confirmation.
 For $\rho^*=0.14$  (part~c) all the approximations, but the HNC1
  predict the density anomaly, whereas at higher bulk density,  $\rho^*=0.2$, (part~d)
  only PY1 and PY2 predict anomalies.

  The range of the densities for which the HNC2 leads to
   anomalous behavior is rather narrow, $0.13 \leqslant \rho^* \leqslant 0.155$.
 As we have shown in our previous work~\cite{ps3}, the RY approximation with the
 global coexistence criterion predicts anomalous behavior up to $\rho^*\approx 0.19$.
 Surprisingly, the PY2
approximation
 yields {\it two} density anomaly regions: the first one is for the densities $[0.07,0.11]$
 and the second one is for
 $\rho^*\geqslant 0.13$. We must recall, however, that the structure predicted from PY2
 approximation  differs very much from the computer simulation results.
 Summary of our calculations is presented in figure~5.  We show here TMD lines resulting
 from different approximate theories. Since the PY2 predictions quantitatively differ
 from the computer simulation results, the relevant curve has been omitted.
 The TMD line from the RY approach evaluated  using   global coexistence
  criterion  differs from that obtained by Barros de Oliveira et al.~\cite{barro2}
  None of the approximate theories is capable of reproducing the simulation
  results at quantitative level.

 Finally, we show the results of the bridge function calculations. The
 bridge function for the model in question has never been  investigated so far.
 The function
 $\gamma(r)$, entering equation~(8) was calculated from
 \begin{equation}
 \gamma(r)= \frac{1}{2\pi^2 r}\int_0^{\infty} \rd k k \sin(kr)\frac{\rho \tilde h(k)^2}
 {1+\rho \tilde h(k)}\,,
 \end{equation}
 where $\tilde h(k)$ is the Fourier transform of the function $h(r)$.
 We should stress that the calculations of the Fourier transform $\tilde h(k)$ and
 the inverse transform in equation~(9) should be carried out with a special
 care, as described by Kolafa et al.~\cite{kolafa1}. Our calculations were performed
 at a constant density equal to $\rho^*=0.152$ for a set of temperatures across the
 TMD line, as well as at a constant temperature, $T^{*}=0.175$ and for several densities,
 again across
 the TMD line (cf. figure~5). The functions $B(r)$ significantly differ from  the functions for
 hard-spheres~\cite{dh1,kolafa1}. In particular, the decay of the bridge functions
 for the core-softened potential is much slower than for hard-spheres.
 Even for $r/\sigma_{\mathrm{f}} \approx 10$, the bridge functions exhibit well pronounced oscillations.
 These oscillations  almost vanish  at
 a  distance as large as $r/\sigma_{\mathrm{f}} \approx 16$, whereas for
 hard-spheres at very high densities, the oscillations already vanish at $r/\sigma_{\mathrm{f}} \approx 5$,
 cf. figure~3 of~\cite{kolafa1}. Crossing the TMD line seems to have no effect on
 the shape of the bridge functions. The evolution of the functions $B(r)$ along
 the temperature and the density branches (figure~6) is smooth and
 we do not observe any peculiarities connected with specific thermodynamic behavior of the system.
 However, the range of oscillations of a bridge function is worth stressing once again.
\begin{figure}[ht]
\begin{center}
\includegraphics[width=0.45\textwidth]{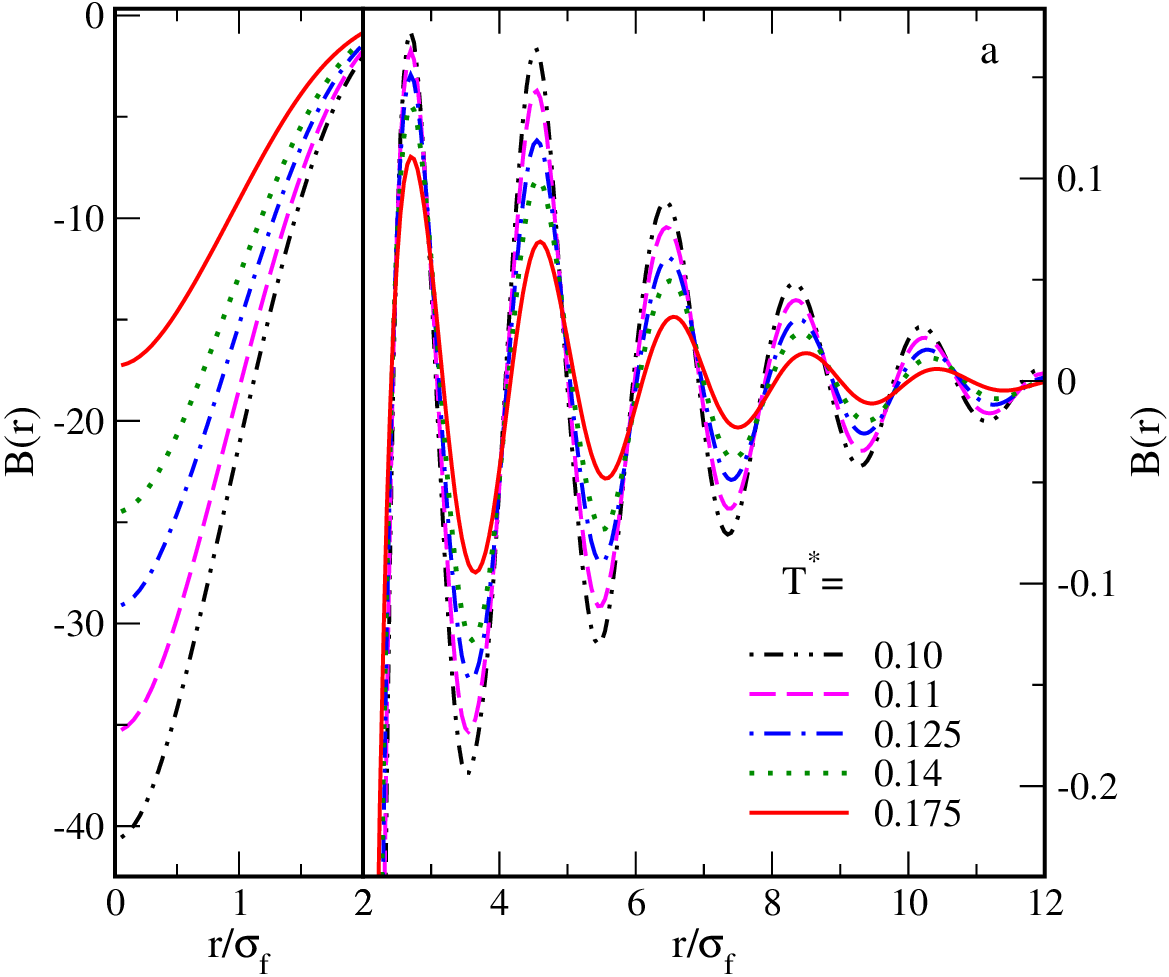}
\includegraphics[width=0.45\textwidth]{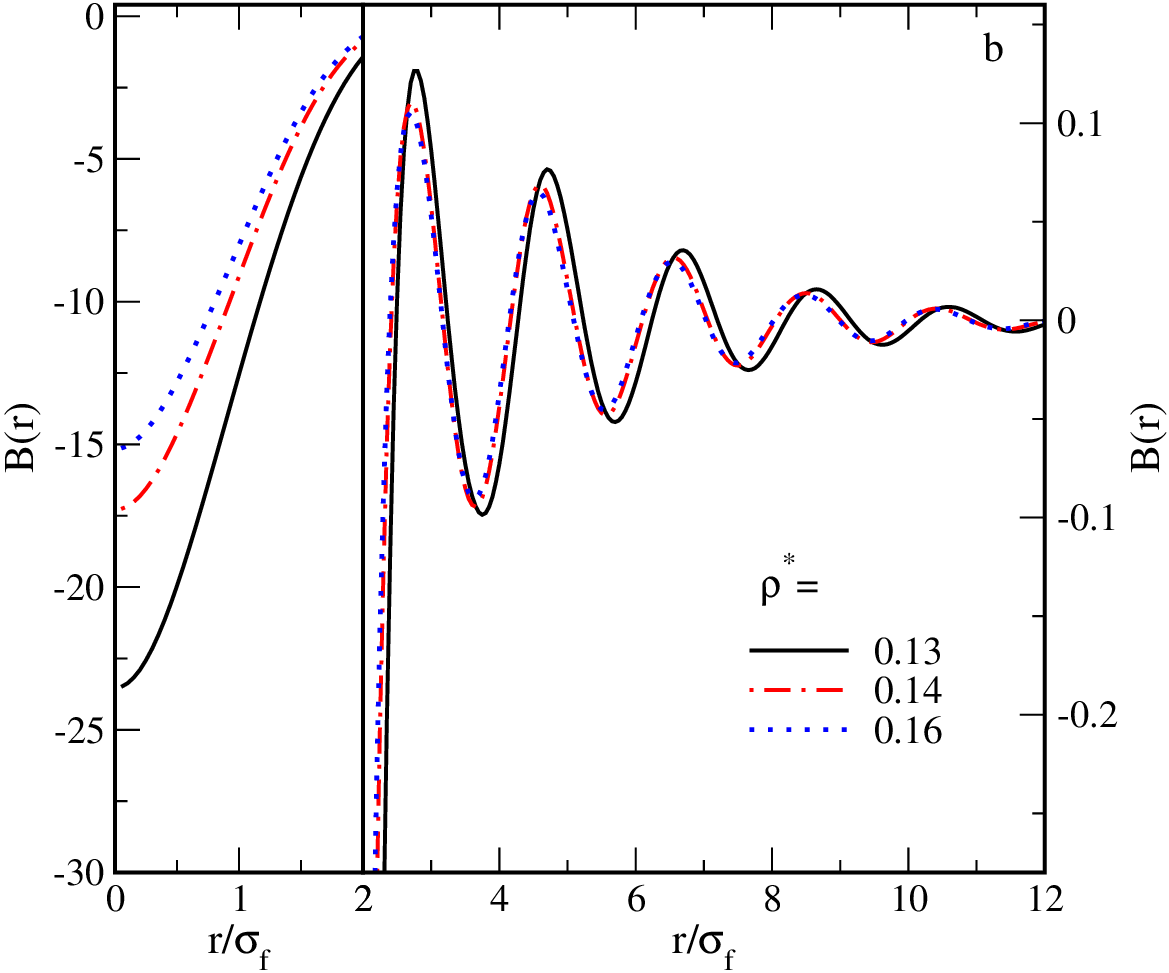}
\end{center}
\caption{
(Color on-line)  Changes of the bridge functions across the TMD
 line. Part~a is at a fixed density, $\rho^*=0.152$, and
 for different temperatures given in the figure,
 whereas part~b is at a fixed temperature, $T^*=0.175$,
 and for three different densities given in
 the figure. The paths along which the calculations were performed
 are shown in the inset to
 figure~5 as dotted lines.}
\end{figure}

 Let us  briefly summarize our findings. We have demonstrated that the
 radial distribution functions predicted by the second-order HNC2 approach
 are more accurate than the predictions of any singlet theory. In contrast to
 the singlet RY approach, the HNC2 approximation does not involve any adjustable
 parameter. However, the  PY2 approximation is inaccurate and the distribution functions
 resulting from it  significantly differ from those of all remaining theories.
 The discrepancies between PY2 and HNC2 approximations are puzzling and
 we cannot offer any physical interpretation of these trends.
 The second-order HNC2 approach yields the density anomaly in the system for
 densities from the interval $[0.13, 0.15]$. The maximum temperature at which
 this phenomenon is observed is $T^*_{\rm max}\approx 0.2$.  Moreover, for the first time we
 have evaluated the bridge functions for the core softened fluid. Unlike the
 hard-sphere or Lennard-Jones fluids, the oscillations of the bridge functions
 extend over much larger interparticle separations.
A successful parametrization of the bridge function similar to hard-sphere systems,
 see e.g.~\cite{kolafa1,grund,LM}, would be helpful in developing new closures for
 singlet level theory for a class of core-softened potentials.

\newpage

\ukrainianpart
\title
{Мікроскопічна структура і термодинаміка  моделі плину з пом'якшеним кором з теорії інтегральних рівнянь другого порядку}

\author{O.~Пізіо\refaddr{label1},
З.~Соколовська\refaddr{label2},
С. Соколовський\refaddr{label3}}

\addresses{\addr{label1}
Інститут хімії УГАМ, Койокан, Мексика
\addr{label2}
Інститут агрофізики Польської академії наук, Люблін, Республіка Польща
\addr{label3}
Університет ім. Марії Складовської-Кюрі, Люблін, Республіка Польща
}

\date{Отримано 18 квітня 2010р., в остаточному вигляді --- 27 липня 2010р.}

\makeukrtitle

\begin{abstract}
\tolerance=3000%
Ми дослідили структурні і термодинамічні властивості однорідного тривимірного плину з пом'як\-шеним кором, використовуючи інтегральні рівняння Орнштейна-Церніке другого порядку із гіперланцюговим замиканням та замиканням Перкуса-Євіка. Зроблено порівняння радіальних функцій розподілу з відповідними функціями, отриманими із синглетних інтегральних рівнянь, а також з даними комп'ютерного моделювання. З різних наближених теорій встановлено границі області  аномалії густини. Отримані результати показують, що гіперланцюгове наближення інтегральних рівнянь другого порядку може бути використане для опису як структури, так і  аномалії густини цього модельного плину. Крім того, ми представляємо результати обчислень місткових функцій.

\keywords теорія рідин, інтегральні рівняння другого порядку,  аномалія густини, місткові функції

\end{abstract}

\end{document}